\documentclass{epl}

\newcommand{\be}{\begin{equation}}
\newcommand{\ee}{\end{equation}}
\newcommand{\bea}{\begin{eqnarray}}
\newcommand{\eea}{\end{eqnarray}}

\newcommand{\aeq}{&=&}

\newcommand{\itDelta}{{\it \Delta}}

\newcommand{\itPi}{{\it \Pi}}

\newcommand{\bra}{\langle}
\newcommand{\ket}{\rangle}
\newcommand{\dbra}{\bra \! \bra}
\newcommand{\dket}{\ket \! \ket}

\newcommand{\me}{\mbox{e}}

\newcommand{\bq}{{\bar q}}

\newcommand{\rL}{{\rm L}}
\newcommand{\rT}{{\rm T}}

\newcommand{\rR}{{\rm R}}

\newcommand{\rd}{{\rm d}}
\newcommand{\rc}{{\rm c}}

\newcommand{\rRe}{{\rm Re}}

\newcommand{\rN}{{\rm N}}
\newcommand{\rS}{{\rm S}}

\title{Analysis of Velocity Derivatives in Turbulence 
based on Generalized Statistics}
\shorttitle{Analysis of Velocity Derivatives in Turbulence}
\author{N. Arimitsu\inst{1} \and T. Arimitsu\inst{2}}
\institute{
  \inst{1} Graduate School of EIS, Yokohama Nat'l.\ University 
  - Kanagawa 240-8501, Japan\\
  \inst{2} Institute of Physics, University of Tsukuba 
  - Ibaraki 305-8571, Japan
}

\begin{document}

\maketitle

\begin{abstract}

A theoretical formula for the probability density function (PDF) 
of {\it velocity derivatives} in a fully developed turbulent flow is derived
with the multifractal aspect based on the generalized 
measures of entropy, i.e., the extensive R\'{e}nyi entropy or the non-extensive 
Tsallis entropy, and is used, successfully, to analyze the PDF's observed
in the direct numerical simulation (DNS) conducted by Gotoh et al..
The minimum length scale $r_{\rd}/\eta$ in the longitudinal (transverse) 
inertial range of the DNS is estimated to be 
$r_{\rd}^{\rm L}/\eta = 1.716$ ($r_{\rd}^{\rm T}/\eta = 2.180$) 
in the unit of the Kolmogorov scale $\eta$. 

%Below this minimum length, the system may dissipate, effectively, its energy.

\end{abstract}

%%%%%%%%%%%%%%%%%%%%%%%%%%%%%%%%%%%%%%%%%%%%%%%%%%%%%%%%%%%%%%%%%%%%%%%%%
%\section{Section title}
%Paper text.
%See fig.~\ref{f.1}, table~\ref{t.1} and eq.~(\ref{e.1}).
%See also~\cite{b.a,b.b}.
%\begin{equation}
%\label{e.1}
%0\neq1
%\end{equation}
%
%\begin{figure}
%\onefigure{epl-template.eps}
%\caption{Figure caption.}
%\label{f.1}
%\end{figure}
%
%\begin{table}
%\caption{Table caption.}
%\label{t.1}
%\begin{center}
%\begin{tabular}{lcr}
%first  & table & row\\
%second & table & row
%\end{tabular}
%\end{center}
%\end{table}
%
%\acknowledgments
%Paper text.
%
%\begin{thebibliography}{0}
%
%\bibitem{b.a}
%  \Name{Author F., Author S. \and Author T.}
%  \REVIEW{Some Rev. A}{69}{1969}{9691}.
%
%\bibitem{b.b}
%  \Name{Author F. \and Author S.}
%  \Book{Some Book of Interest}
%  \Editor{A. Editor}
%  \Vol{9}
%  \Publ{Publishing house, City}
%  \Year{1939}
%  \Page{666}.
%
%\bibitem{b.c}
%  \Editor{Editor A.}
%  \Book{Some Book of Interest}
%  \Vol{9}
%  \Publ{Publishing house, City}
%  \Year{1939}
%  \Page{666}.
%
%\end{thebibliography}
%%%%%%%%%%%%%%%%%%%%%%%%%%%%%%%%%%%%%%%%%%%%%%%%%%%%%%%

%%%%%%%%%%%%%%%%%%%%%%%%%%%%%%%%%%%%%%%%%%%%%%%%%%%%%%%%%%%%%%%%%%%
% Text
%%%%%%%%%%%%%%%%%%%%%%%%%%%%%%%%%%%%%%%%%%%%%%%%%%%%%%%%%%%%%%%%%%%

%\section{Introduction}

In the previous paper~\cite{AA7}, 
we analyzed, precisely, the probability density functions (PDF's) 
of velocity fluctuations observed in 
the direct numerical simulation (DNS) of turbulence conducted 
by Gotoh et al.~\cite{Gotoh02} at the Taylor microscale Reynolds number 
$R_\lambda = 381$.
At this Reynolds number, the PDF's had been measured with high accuracy up to 
the order of $10^{-9} \sim 10^{-10}$.
The correctness of the analytical formula 
for the scaling exponents of velocity structure function~\cite{AA1,AA2,AA3,AA4} 
enabled us to extract the value of the intermittency exponent $\mu$ by fitting
it with the ten observed data in the DNS by the method of least square~\cite{AA7}.
With the intermittency exponent, the parameters in the analytical formula of 
the PDF of velocity fluctuations~\cite{AA4,AA5,AA6} are determined, self-consistently.
The formula was used to fit each of the observed PDF's of velocity fluctuations 
for ten different separations $r/\eta$ by means of
the method of least square~\cite{AA7}, and to extract, successfully, the dependence
of the number $n$ of steps in the energy cascade on the separation $r/\eta$:
\bea
n \aeq -1.050 \times \log_2 r/\eta + 16.74
\quad (\mbox{for } \ell_{\rc}^{\rm L} \leq r),
\label{n-roeta L larger} \\
n \aeq -2.540 \times \log_2 r/\eta + 25.08
\quad (\mbox{for } r < \ell_{\rc}^{\rm L})
\label{n-roeta L less}
\eea
with the crossover length $\ell_{\rc}^{\rm L}/\eta = 48.26$ 
for {\it longitudinal} fluctuations, and
\bea
n \aeq -0.9896 \times \log_2 r/\eta + 13.95
\quad (\mbox{for } \ell_{\rc}^{\rm T} \leq r),
\label{n-roeta T larger} \\
n \aeq -2.820 \times \log_2 r/\eta + 23.87
\quad (\mbox{for } r < \ell_{\rc}^{\rm T})
\label{n-roeta T less}
\eea
with the crossover length $\ell_{\rc}^{\rm T}/\eta = 42.57$ 
for {\it transverse} fluctuations~\footnote{
Here, $\eta$ is the Kolmogorov scale~\cite{K41} defined by 
$
\eta = ( \nu^3/\epsilon )^{1/4}
$
with $\nu$ being the kinematic viscosity, and $\epsilon$ ($=\epsilon_0$) 
the energy input rate to the largest eddies with size $\ell_0$, 
and has the value $\eta = 2.58 \times 10^{-3}$ in the DNS~\cite{Gotoh02} 
at $R_\lambda = 381$.
}.
These straight lines on semi-logarithmic sheet, explicitly, told us
that there exist two scaling regions~\cite{AA7}, i.e., 
the {\it upper} scaling region with larger separations which may correspond 
to the scaling range observed by Gotoh et al.~\cite{Gotoh02},
and the {\it lower} scaling region with smaller separations which is 
another scaling region extracted first by the systematic analyses in \cite{AA7}.
These scaling regions are divided by the crossover lengths 
$\ell_{\rc}^{\rm L}/\eta$ and $\ell_{\rc}^{\rm T}/\eta$
approximately of the order of the Taylor microscale 
$\lambda /\eta = 38.33$ reported in \cite{Gotoh02} at $R_\lambda = 381$.

We can view that the turbulent flow, satisfying the Navier-Stokes equation
\be
\partial {\vec u}/\partial t
+ ( {\vec u}\cdot {\vec \nabla} ) {\vec u} 
= - {\vec \nabla} \left(p/\rho \right)
+ \nu \nabla^2 {\vec u}
\label{N-S eq}
\ee
of an incompressible fluid,
consists of a cascade of eddies 
with different diameter of the order of $\ell_n = \delta_n \ell_0$ 
where $\delta_n = 2^{-n}$ $(n=0,1,2,\cdots)$.
The quantities $\rho$ and $p$ represent, respectively, the mass density and 
the pressure.
At each step of the cascade, say at the $n$th step, mother-eddies of size $\ell_{n-1}$
produce smaller daugter-eddies having a half of mother's diameter 
with the energy-transfer rate
$\epsilon_n$ that represents the rate of transfer of energy per unit mass 
from eddies of size $\ell_{n-1}$ to those of size $\ell_n$
(the energy cascade model).
Following formally the energy cascade model, we are measuring space scale 
by $\ell_n$. However our analysis in the following is not restricted within 
the energy cascade model, i.e., the number of step $n$ can be real number.

In this paper, we will derive the formula for the PDF of {\it velocity derivatives}
in fully developed turbulence by the statistics based on the generalized entropy, 
i.e., R\'{e}nyi's \cite{Renyi} or Tsallis' \cite{Tsallis88,Tsallis99},
and will analyze the velocity derivative PDF's obtained in 
the DNS~\cite{Gotoh02} at $R_\lambda = 381$ having far better accuracy
than that in any previous experiments, real or numerical.
The main interest, here, is the {\it longitudinal} velocity derivative 
$
\partial u({\vec r}) / \partial r_1
$
and the {\it transverse} velocity derivative
$
\partial u({\vec r}) / \partial r_2
$ 
(or 
$
\partial u({\vec r}) / \partial r_3
$),
where ${\vec r}=(r_1,r_2,r_3)$, and $u$ is the $r_1$-component of 
the fluid velocity field ${\vec u}$
of the turbulent flow produced by a grid with size $\ell_0$
putting in a laminar flow parallel to the $r_1$ direction.
Introducing the velocity difference $\delta u_n$ of the component $u$ 
at two points separated by the distance $\ell_n$, the velocity derivatives
may be estimated by 
$
\lim_{\ell_n \rightarrow 0} \delta u_n / \ell_n
$ 
($
= \lim_{n \rightarrow \infty} \delta u_n / \ell_n
$).
The Reynolds number $\rRe$ of the system is given by 
$
{\rm Re} = \delta u_0 \ell_0/\nu = ( \ell_0/\eta )^{4/3}
$.
For high Reynolds number $\rRe \gg 1$, or for the situation where 
effects of the kinematic viscosity $\nu$ can be neglected compared with
those of the turbulent viscosity, the Navier-Stokes equation (\ref{N-S eq})
is invariant under 
the scale transformation~\cite{Frisch-Parisi83,Meneveau87b}:
${\vec r} \rightarrow \lambda {\vec r}$, 
${\vec u} \rightarrow \lambda^{\alpha/3} {\vec u}$, 
$t \rightarrow \lambda^{1- \alpha/3} t$ and 
$\left(p/\rho\right) \rightarrow \lambda^{2\alpha/3} \left(p/\rho\right)$.
The exponent $\alpha$ is an arbitrary real quantity which specifies the degree
of singularity in the velocity derivative~\cite{Benzi84}
for $\alpha < 3$, i.e.,
$
\lim_{\ell_n \rightarrow 0} \delta u_n/\ell_n
\sim \lim_{\ell_n \rightarrow 0} \ell_n^{\alpha/3-1}
$
which can be seen with the relation
$
\delta u_n / \delta u_0 = (\ell_n / \ell_0)^{\alpha/3}
\label{u-alpha}
$.

The energy $E_n = (\delta u_n)^2/2$ per unit mass contained 
in an eddy of size $\ell_n$ is estimated as $E_n \sim (\ell_n \omega_n)^2$
where $\omega_n \sim \delta u_n / \ell_n$ represents 
the angular momentum of the eddy.
We see that $\lim_{\ell_n \rightarrow 0} \omega_n$ has the same singularity
as the velocity derivative does.
Within the region satisfying the scale invariance, we have
$E_n = E_0 \delta_n^{2\alpha/3}$.
The energy spectrum $E(k)$, defined through
$E_n = \int_{k_n}^{k_{n+1}} dk E(k)$ with $k_n = \ell_n^{-1}$, 
has the wavenumber dependence 
$E(k) \propto k^{-1-2\alpha/3}$.
By the way, the energy transfer rate, estimated by
$\epsilon_n \sim E_n \omega_n$, satisfies
$\epsilon_n = \epsilon_0 \delta_n^{\alpha -1}$ 
in the region of the scale invariance.
Kolmogorov's assumption in K41 \cite{K41} that
there is no fluctuation in $\epsilon_n$ leads us to $\alpha = 1$.
With this value of $\alpha$, $E(k)$ represents the Kolmogorov energy spectrum,
i.e., $E(k) \propto k^{-3/5}$.
In order to explain the intermittency in turbulence, we shall introduce
a fluctuation in $\alpha$.

The present analysis rests on the assumption that the distribution of
the exponent $\alpha$ is multifractal, and that the probability 
$
P^{(n)}(\alpha) d\alpha
$
to find, at a point in physical space, an eddy of size $\ell_n$ having
a value of the degree of singularity in the range 
$
\alpha \sim \alpha + d \alpha
$
is given by
$
P^{(n)}(\alpha) = [P^{(1)}(\alpha)]^n
$
with \cite{AA1,AA2,AA3,AA4}
\be
P^{(1)}(\alpha) \propto \left[ 1 - (\alpha - \alpha_0)^2 \big/ (\itDelta \alpha )^2 
\right]^{1/(1-q)}, \quad
(\itDelta \alpha)^2 = 2X \big/ [(1-q) \ln 2 ].
\label{Tsallis prob density}
\ee
Here, it is assumed that each step in the cascade is statistically independent.
The distribution function (\ref{Tsallis prob density}) is 
derived by taking an extremum of the generalized entropy~\footnote{
The R\'{e}nyi entropy 
$
S_{q}^{\rR}[P^{(1)}(\alpha)] = \left(1-q \right)^{-1} 
\ln \int d \alpha P^{(1)}(\alpha)^{q}
\label{SqR-alpha}
$~\cite{Renyi}
has the extensive character as the usual thermodynamical entropy does,
whereas the Tsallis entropy
$
S_{q}^{\rT}[P^{(1)}(\alpha)] = \left(1-q \right)^{-1}
\left(\int d\alpha \ P^{(1)}(\alpha)^{q} -1 \right)
\label{SqTHC-alpha}
$~\cite{Tsallis88,Tsallis99,Havrda-Charvat}
is non-extensive.
In spite of different characteristics of these entropies
the distribution functions giving the extremum of each entropy
have the common structure (\ref{Tsallis prob density}).
}
with the two constraints, i.e., the normalization of distribution function:
$
\int d\alpha P^{(1)}(\alpha) = \mbox{const.}
\label{cons of prob}
$
and the $q$-variance being kept constant as a known quantity:
$
\sigma_q^2 = (\int d\alpha P^{(1)}(\alpha)^{q} 
(\alpha- \alpha_0 )^2 ) / \int d\alpha P^{(1)}(\alpha)^{q}
\label{q-variance}
$.

The dependence of the parameters $\alpha_0$, $X$ and $q$ on 
the intermittency exponent $\mu$ is determined, 
self-consistently, with the help of the three independent equations, i.e.,
the energy conservation:
$
\left\bra \epsilon_n \right\ket = \epsilon
\label{cons of energy}
$,
the definition of the intermittency exponent $\mu$:
$
\bra \epsilon_n^2 \ket 
= \epsilon^2 \delta_n^{-\mu}
\label{def of mu}
$,
and the scaling relation~\footnote{
The scaling relation is a generalization of the one derived first in
\cite{Costa,Lyra98} to the case where the multifractal spectrum
has negative values.
}:
$
1/(1-q) = 1/\alpha_- - 1/\alpha_+
\label{scaling relation}
$
with $\alpha_\pm$ satisfying $f(\alpha_\pm) =0$ 
where the multifractal spectrum~\cite{AA1,AA2,AA3,AA4}
\be
f(\alpha) = 1 + (1-q)^{-1} \log_2 \left[ 1 - \left(\alpha - \alpha_0\right)^2
\big/ \left(\Delta \alpha \right)^2 \right]
\label{Tsallis f-alpha}
\ee
is derived by the relation
$
P^{(n)}(\alpha) \propto \delta_n^{1-f(\alpha)}
$ \cite{Meneveau87b,AA4}
that reveals how densely each singularity, labeled by $\alpha$, fills physical space.
The average $\bra \cdots \ket$ is taken with $P^{(n)}(\alpha)$.
Note that the relation between $\alpha$ and $\epsilon_n$ is given by
$
\epsilon_n/\epsilon = \left(\ell_n/\ell_0 \right)^{\alpha -1}
\label{u/epsilon-alpha}
$.
For the region where the value of $\mu$ is usually observed,
i.e., $0.13 \leq \mu \leq 0.40$,
the three self-consistent equations are solved to give 
the approximate equations~\cite{AA7}:
$
\alpha_0 = 0.9989 + 0.5814 \mu
$,
$
X = - 2.848 \times 10^{-3} + 1.198 \mu
$
and
$
q = -1.507 + 20.58 \mu - 97.11 \mu^2 + 260.4 \mu^3 - 365.4 \mu^4 + 208.3 \mu^5
$.

Let us suppose that $\ell_{\rd}$ is the typical length giving the minimum scale
within the energy cascade model. With this shortest length $\ell_{\rd}$, 
the velocity derivatives may be given by
$
\vert s \vert = \delta u_{\rd}/\ell_{\rd}
$
with 
$
\delta u_{\rd} = \delta u_{n=n_{\rd}}
$
where $n_{\rd}$ is introduced through
$
\ell_{\rd}/\ell_0 = \delta_{\rd} = 2^{-n_{\rd}}
$.
Since there are two mechanisms in turbulent flow to rule 
its dissipative evolution, i.e., the one controlled by the kinematic viscosity 
that takes care thermal fluctuations, and the other by the turbulent viscosity
that is responsible for intermittent fluctuations related to the singularities 
in velocity derivative, it may be reasonable to assume that the probability 
$\itPi(x) dx$ to find the scaled velocity derivative
$
x = \delta_{\rd} t_0 s 
$
in the range $x \sim x+dx$ can be divided into two parts:
\be
\itPi(x) dx = \itPi_{\rS}(\vert x \vert) dx
+ \Delta \itPi(x) dx.
\ee
Here, the singular part PDF $\itPi_{\rS}(\vert x \vert)$ represents
the contribution from multifractal distribution of the singularities,
and the correction part $\Delta \itPi(x)$ from the viscus term neglected
in the scale transformation.
The former is derived through
$
\itPi_{\rS}(\vert x \vert) dx = P^{(n)}(\alpha) d \alpha
$
with the transformation of the variables:
$
\vert x \vert = \delta_{\rd}^{\alpha/3}
$.
The $m$th moments of the velocity derivatives, defined by
$
\dbra \vert x \vert^m \dket 
= \int_{-\infty}^{\infty} dx  
\vert x \vert^m \itPi(x)
$,
are given by 
\be
\dbra \vert x \vert^m \dket = 2 \gamma_m
+ (1-2\gamma_0 ) \
a_m \ \delta_{\rd}^{\zeta_m}
\ee
with
$
a_{3\bq} = \{ 2 / [C_{\bq}^{1/2} ( 1+ C_{\bq}^{1/2} ) ] \}^{1/2}
$,
$
{C}_{\bq}= 1 + 2 \bq^2 (1-q) X \ln 2
\label{cal D}
$
and
$
2\gamma_m = \int_{-\infty}^{\infty} dx\ 
\vert x \vert^m \itPi_\rN(x)
$.
We used the normalization: $\dbra 1 \dket = 1$.
The quantity 
\be
\zeta_m = \alpha_0 m /3 
- 2Xm^2 \big/\left[9 \left(1+{C}_{m/3}^{1/2} \right) \right] 
- \left[1-\log_2 \left(1+{C}_{m/3}^{1/2} \right) \right] 
\big/(1-q) 
\label{zeta}
\ee
is the so-called scaling exponent of the velocity structure function,
whose expression was derived first by the present authors \cite{AA1,AA2,AA3,AA4}.
Note that the formula is independent of the length $\ell_{\rd}$, and, therefore, 
independent of $\ell_n$.

\begin{table}[htbp]
\begin{center}
\begin{tabular}{ccccccc}
  & \multicolumn{2}{c}{longitudinal} & \multicolumn{2}{c}{transverse} \\
\hline
$\mu$      & \multicolumn{2}{c}{0.240}  & \multicolumn{2}{c}{0.327} \\
\hline
$q$        & \multicolumn{2}{c}{0.391}  & \multicolumn{2}{c}{0.543} \\
$\alpha_0$ & \multicolumn{2}{c}{1.138}  & \multicolumn{2}{c}{1.189} \\
$X$        & \multicolumn{2}{c}{0.285}  & \multicolumn{2}{c}{0.388} 
%\hline\hline
\end{tabular}
\end{center}
\caption{Values of the intermittency exponent $\mu$ and the parameters
$q$, $\alpha_0$ and $X$ for longitudinal and transverse velocity fluctuations
\cite{AA7}, determined by the formula (\ref{zeta}) being consistent with 
the observed data in DNS conducted by Gotoh et al..
}
\label{parameters}
\end{table}

With the help of the analytical formula (\ref{zeta}), we determined in \cite{AA7} 
the values of the intermittency exponent $\mu$
and the parameters $q$, $\alpha_0$ and $X$ 
by fitting the ten DNS data of the scaling exponents $\zeta_m$ 
($m = 1,2,\cdots, 10$) at $\rR_\lambda = 381$ \cite{Gotoh02} 
with the method of least squares.
The determined values are listed in table~\ref{parameters} both for 
the longitudinal and transverse velocity fluctuations.
Note that the relation $\mu = 2 - \zeta_6$ is satisfied within 
the experimental error bars. 
We have
$
\alpha_{+} -\alpha_0 
= \alpha_0 - \alpha_{-} = 0.6818
$ (0.8167),
$
\itDelta \alpha = 1.160
$ (1.566)
for the longitudinal (transverse) fluctuations.

Since we are interested in the large deviation stemmed from the singular part
$\itPi_{\rS}(\vert x \vert)$ that may contribute to the symmetric part of the PDF,
we will symmetrize right and left of the experimental PDF,
and will compare it with our theoretical PDF in the following.

Let us introduce 
the PDF $\hat{\itPi}(\vert \xi \vert)$ of the velocity derivatives by
$
\hat{\itPi}(\vert \xi \vert) d\xi 
= \itPi(\vert x \vert) dx
$
with the new variable 
\be
\xi = s / \dbra s^2 \dket^{1/2} 
= x / \dbra x^2 \dket^{1/2}, \quad \vert \xi \vert = \bar{\xi} \delta_{\rd}^{\alpha /3 -\zeta_2 /2},
\ee
scaled by the variance of velocity derivatives.
It may be appropriate to devide the PDF into two parts:
\bea
\hat{\itPi}(\xi) \aeq \hat{\itPi}_{*<}(\xi)
\quad \mbox{for $\xi^* \leq \vert \xi \vert \leq 
\bar{\xi} \delta_{\rd}^{\alpha_{\rm min} /3 -\zeta_2 /2}$}
\label{PDF der1}
\\
\hat{\itPi}(\xi) \aeq \hat{\itPi}_{<*}(\xi)
\quad \mbox{for $\vert \xi \vert \leq \xi^*$}
\label{PDF der2} 
\eea
with the point $\xi^*$ defined by 
$
\xi^* = \bar{\xi} \delta_{\rd}^{\alpha^* /3 -\zeta_2 /2}
$
where $\alpha^*$ is the solution of 
$
\zeta_2/2 -\alpha/3 +1 -f(\alpha) = 0
$ that provides us with the least $n_{\rd}$-dependence of $\hat{\itPi}(\xi^*)$.
Here, 
$
\bar{\xi} = [2 \gamma_2 \delta_{\rd}^{-\zeta_2} + (1-2\gamma_0 ) 
a_2 ]^{-1/2}
$.
As the value of $\xi^*$ turns out to be of order 1 (see below), 
we are deviding the region by the order of the variance of
velocity derivative.
$\hat{\itPi}_{<*}(\xi)$ and $\hat{\itPi}_{*<}(\xi)$ 
are connected at $\xi^*$ under the condition that they should have 
the same value and the same slope there.
In our analysis, since we are assuming that the large deviations of 
the velocity derivative come from the maltifractal distribution of
these singularities in real space \cite{Benzi84},
it is consistent to put
$
\hat{\itPi}_{*<}(\xi) = \hat{\itPi}_{\rS}(\vert \xi \vert)
$.
This leads
\be
\hat{\itPi}_{*<}(\xi)
= \bar{\itPi}_{\rS} \left( \bar{\xi} \big/\vert \xi \vert\right)
\left\{1 - \left[3 \ln \left\vert \xi / \xi_0 \right\vert \big/
\left(\Delta \alpha \ \vert \ln \delta_{\rd} \vert \right) \right]^2
\right\}^{n_{\rd}/(1-q)}
\label{PDF larger}
\ee
with 
$
\vert \xi_0 \vert = \bar{\xi} \delta_{\rd}^{\alpha_0 /3 -\zeta_2 /2}
$.
For smaller velocity derivatives $\vert \xi \vert \le \xi^*$, 
the contribution to the PDF may come, mainly, from
the term ignored in the scale transformation producing, i.e.,
the term responsible for thermal fluctuations related to the kinematic viscosity, 
and also from measuremant errors,
we take for the PDF $\hat{\itPi}_{<*}(\xi)$ a Gaussian function.
The connection at $\xi^*$ with (\ref{PDF larger}) gives us 
%the specific Gaussian form
\be
\hat{\itPi}_{<*}(\xi) = \bar{\itPi}_{\rS}
\me^{-[1+3f'(\alpha^*)] [(\xi/\xi^* )^2 -1 ] /2}
\label{PDF less}
\ee
with
$
\bar{\itPi}_{\rS} = 3 (1-2\gamma_0)
/ (2 \bar{\xi} \sqrt{2\pi X \vert \ln \delta_{\rd} \vert} )
$.
It is remarkable that the PDF $\hat{\itPi}(\vert \xi \vert)$ 
of the {\it velocity derivative}, given by (\ref{PDF der1}) and (\ref{PDF der2})
with (\ref{PDF larger}) and (\ref{PDF less}), 
turns out to have the same structure as the PDF $\hat{\itPi}^{(n)}(\vert \xi_n \vert)$ 
of the {\it velocity fluctuations} \cite{AA4,AA5,AA6,AA7} with the separation 
$\ell_n = \ell_{\rd}$.

\begin{figure}[htbq]
\onefigure[width=14cm,clip]{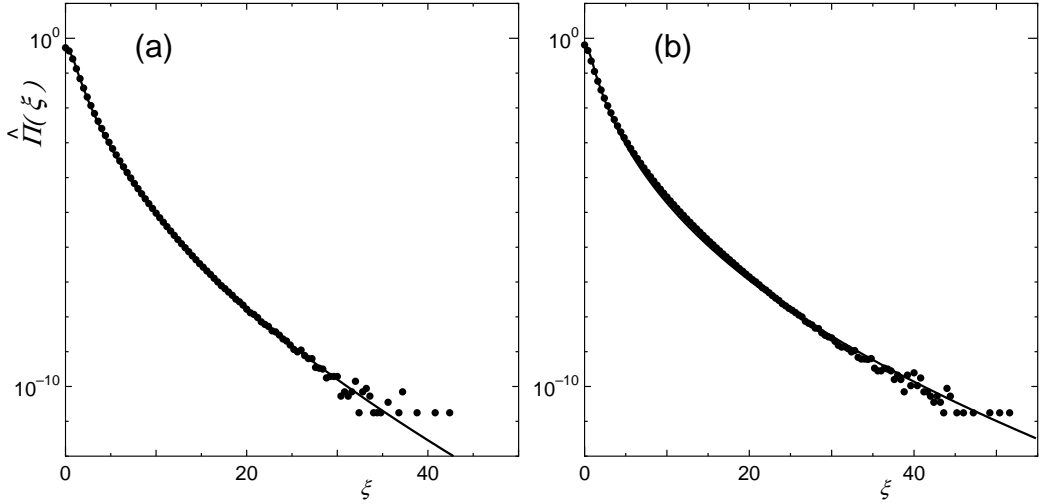}
\caption{PDF's of the velocity derivatives for (a) longitudinal and for 
(b) transverse. Closed circles are the symmetrized points of 
the PDF observed in DNS conducted by Gotoh et al.\ 
at $R_\lambda = 381$~\cite{Gotoh02}.
Solid line represents the PDF given by the analytic formula 
(\ref{PDF der1}) and (\ref{PDF der2}) with the parameters given 
in Table~\ref{parameters}, and with (a) $n_{\rd}= n_{\rd}^{\rL}=23.1$ and
(b) $n_{\rd}= n_{\rd}^{\rT}=20.7$. }
\label{vel grad}
\end{figure}

%\begin{figure}
%\onefigure[width=7cm,clip]{puyb11.eps}
%\caption{Figure caption.}
%\label{vel grad transverse}
%\end{figure}

The comparison between the PDF's of the velocity derivatives
at $R_\lambda = 381$ measured in the DNS \cite{Gotoh02} and 
those obtained by the present analysis is given in fig.~\ref{vel grad}~(a)
for longitudinal derivative and in fig.~\ref{vel grad}~(b) 
for transverse derivative.
In order to extract the symmetrical part of the PDF, we took
mean average between the DNS data on the left hand side and that on the right hand side.
The symmetrized data are described by closed circles.
The solid lines are the curves of $\hat{\itPi}(\xi)$ given by 
(\ref{PDF larger}) and (\ref{PDF less}) with 
the values of parameters in table~\ref{parameters}.
Note that $\xi^* = 0.982$ for the longitudinal velocity derivative,
whereas $\xi^* = 0.900$ for the transverse derivative.
The number $n_{\rd}^{\rm L} = 23.1$ ($n_{\rd}^{\rm T} = 20.7$) of steps 
in the cascade for 
longitudinal (transverse) velocity fluctuations is derived by the method of 
least squares with respect to the logarithm of PDF's for the best fit of 
our theoretical formulae, consisting of (\ref{PDF larger}) and (\ref{PDF less}), 
to the observed values of the PDF by discarding those points which have 
observed values less than 10$^{-10}$ since they scatter largely 
in the logarithmic scale. 
We see an excellent agreement between the measured PDF for the velocity 
derivatives and the analytical formula of PDF derived by 
the present self-consistent theory.

Substituting the obtained values $n_{\rd}^{\rm L} = 23.1$ ($n_{\rd}^{\rm T} = 20.7$) of
the number of steps in the energy cascade for longitudinal (transverse) 
velocity fluctuations into (\ref{n-roeta L less}) (into (\ref{n-roeta T less})),
we obtain the shortest length $r_{\rd}$ of separation in the inertial range 
with the value $r_{\rd}^{\rm L}/\eta = 1.716$ ($r_{\rd}^{\rm T}/\eta = 2.180$).
Then, we can conclude that the range of the {\it lower} scaling region 
in the inertial range of the longitudinal (transverse) velocity fluctuations
is given by $r_{\rd}^{\rm L}/\eta \leq r/\eta \leq \ell_{\rc}^{\rm L}/\eta$ 
($r_{\rd}^{\rm T}/\eta \leq r/\eta \leq \ell_{\rc}^{\rm T}/\eta$).
Adding to this the {\it upper} scaling region, the total inertial range
for longitudinal (transverse) fluctuations within the DNS~\cite{Gotoh02} 
at $R_\lambda = 381$ turns out to be
$1.716 \leq r/\eta \leq 1220$ ($2.180 \leq r/\eta \leq 1220$), where the value
1220 is the largest separation taken by Gotoh et al.~\cite{Gotoh02}
for the measurement of the PDF's of velocity fluctuations.
Note that the largest scale $\ell_0$ for the longitudinal (transverse) fluctuations
can be estimated by putting $n=0$
into (\ref{n-roeta L larger}) (into (\ref{n-roeta T larger})), i.e., 
$\ell_0^{\rm L}/\eta = 6.299 \times 10^4$ ($\ell_0^{\rm T}/\eta = 1.752 \times 10^4$)
leading to $\rRe^{\rm L}=2.506 \times 10^6$ ($\rRe^{\rm T}=4.550 \times 10^5$).
We see that the shortest length scale $\ell_{\rd}/\eta= 2^{-n_{\rd}} \ell_0/\eta$
for longitudinal (transverse) velocity fluctuations within the energy cascade model 
becomes $\ell_{\rd}^{\rm L}/\eta = 7.006 \times 10^{-3}$ 
($\ell_{\rd}^{\rm T}/\eta = 1.029 \times 10^{-2}$) which is different 
from the length $r_{\rd}^{\rm L}/\eta$ ($r_{\rd}^{\rm T}/\eta$) 
giving the estimate of the lowest end of inertial range in the DNS~\cite{AA7} at
$R_\lambda = 381$.
We presume, here, that $r/\eta$ in the formulae 
(\ref{n-roeta L larger}), (\ref{n-roeta L less}) and 
(\ref{n-roeta T larger}), (\ref{n-roeta T less}) 
provides us with a {\it real} distance in the support of the velocity fields
${\vec u}({\vec r})$.

Summarizing, we derived in this paper the formula for PDF of velocity derivatives,
(\ref{PDF der1}) and (\ref{PDF der2}) with (\ref{PDF larger}) and (\ref{PDF less}),
and showed that it explains the observed PDF's in the DNS~\cite{Gotoh02}, 
precisely up to the order of $10^{-10}$, 
with the parameters in Table~\ref{parameters}.
%determined self-consistently
%by making use of the ten reported values of the scaling exponents in the same DNS.
The latter analysis provides us with the number $n_{\rd}$ of steps 
in the energy cascade.
The shortest length scale $r_{\rd}^{\rm L}$ ($r_{\rd}^{\rm T}$), 
which serves the lowest end of the inertial range, is derived by making use of
the obtained value of $n_{\rd}^{\rm L}$ ($n_{\rd}^{\rm T}$) by assuming that 
the formula (\ref{n-roeta L less}) (the formula (\ref{n-roeta T less})) 
between $n$ and $r/\eta$, derived in the analyses of the PDF of longitudinal 
(transverse) velocity fluctuations~\cite{AA7}, is applicable even for 
this shorter scale of length.

Let us close this paper by mentioning something about another trial
for deriving the PDF of velocity derivatives.
We introduced, as a preliminary test, the same philosophy for the definition of 
velocity derivative proposed by Benzi et al.~\cite{Benzi91grad}
into the present analysis based on the generalized statistics, and saw
that the PDF of velocity derivatives thus obtained
(see (43) in \cite{AA4}) cannot explain the PDF's provided 
in the DNS~\cite{Gotoh02} so precise as the present PDF given by
(\ref{PDF der1}) and (\ref{PDF der2}).
A further investigation on the relation between the approach in 
\cite{Benzi91grad} and the present one may be one of the interesting
future problems.

The authors are grateful to Prof.~T.~Gotoh for enlightening discussion
and his kindness to show his data prior to publication.
The authors would like to thank Prof.~C.~Tsallis 
for his fruitful comments with encouragement.

%%%%%%%%%%%%%%%%%%%%%%%%%%%%%%%%%%%%%%%%%%%%%%%%%%%%%%%%%%%%%%%%%
% References:
%%%%%%%%%%%%%%%%%%%%%%%%%%%%%%%%%%%%%%%%%%%%%%%%%%%%%%%%%%%%%%%%%


\begin{thebibliography}{0}

\bibitem{AA7} T.~Arimitsu and N.~Arimitsu, J.~Phys.: Condens.~Matter, 
		{\bf 14}, 2237 (2002).
\bibitem{Gotoh02} T.~Gotoh, D.~Fukayama and T.~Nakano, Phys. Fluids, 
		in press (2002).
\bibitem{AA1} T.~Arimitsu and N.~Arimitsu, J. Phys. A: Math. Gen. {\bf 33}, 
		L235 (2000)  [{\footnotesize CORRIGENDUM}: {\bf 34}, 673 (2001)].
\bibitem{AA2} T.~Arimitsu and N.~Arimitsu, Chaos, Solitons and Fractals 
		{\bf 13}, 479 (2002).
\bibitem{AA3} T.~Arimitsu and N.~Arimitsu, Prog.~Theor.~Phys. {\bf 105}, 
		 355 (2001).
\bibitem{AA4} T.~Arimitsu and N.~Arimitsu, Physica A {\bf 295}, 177 (2001).
\bibitem{AA5} N.~Arimitsu and T.~Arimitsu, J. Korean Phys.\ Soc.\ {\bf 40}, 1032 (2002).
\bibitem{AA6} T.~Arimitsu and N.~Arimitsu, Physica A {\bf 305}, 218 (2002).
\bibitem{K41} A.N.~Kolmogorov, 
		C.R.~Acad. Sci. USSR {\bf 30}, 301; 538 (1941).
\bibitem{Renyi} A.~R\'{e}nyi, {\it Proc.\ 4th Berkeley Symp.\ 
		Maths.\ Stat.\ Prob.} {\bf 1}, 547 (1961).
\bibitem{Tsallis88} C.~Tsallis, J. Stat. Phys. {\bf 52}, 479 (1988).
\bibitem{Tsallis99} C.~Tsallis, Braz. J. Phys. {\bf 29}, 1 (1999);
	On the related recent progresses see at\\ http://tsallis.cat.cbpf.br/biblio.htm.
\bibitem{Meneveau87b} C.~Meneveau and K.R.~Sreenivasan,
		Nucl. Phys. B (Proc. Suppl.) {\bf 2}, 49 (1987).
\bibitem{Frisch-Parisi83} U.~Frisch and G.~Parisi, in {\it Turbulence
		and Predictability in Geophysical Fluid Dynamics and Climate 
		Dynamics}, ed.\ by M.~Ghil, R.~Benzi and G.~Parisi (North-Holland,
		New York, 1985) 84.
\bibitem{Benzi84} R.~Benzi, G.~Paladin, G.~Parisi and A.~Vulpiani, 
		J. Phys. A: Math. Gen. {\bf 17}, 3521 (1984).
\bibitem{Havrda-Charvat} J.H.~Havrda and F.~Charvat,
		Kybernatica {\bf 3}, 30 (1967).
\bibitem{Costa} U.M.S.~Costa, M.L.~Lyra, A.R.~Plastino and
		C.~Tsallis, Phys. Rev. E {\bf 56}, 245 (1997).
\bibitem{Lyra98} M.L.~Lyra and C.~Tsallis, 
		Phys. Rev. Lett. {\bf 80}, 53 (1998).
\bibitem{Benzi91grad} R.~Benzi, L.~Biferale, G.~Paladin, A.~Vulpiani and 
		M.~Vergassola, Phys. Rev. Lett. {\bf 67}, 2299 (1991).

\end{thebibliography}
\end{document}